\documentclass[reprint, aps, prl, longbibliography]{revtex4-1}
\usepackage{graphicx}
\usepackage{color}
\begin{document}
\title{Energy Cascade in Quantum Gases}
\author{Tin-Lun Ho and X. Y. Yin}
\affiliation{Department of Physics, The Ohio State University, Columbus, OH 43210, USA}
\date{\today}

\begin{abstract}
Energy cascade is ubiquitous in systems far from equilibrium. Facilitated by particle interactions and external forces, it can lead to highly complex phenomena like fully developed turbulence, characterized by power law velocity correlation functions. Yet despite decades of research, how these power laws emerge from first principle remains unclear. Recently, experiments show that when a Bose condensate is subjected to periodic shaking, its momentum distribution exhibits a power law behavior. The flexibility of cold atom experiments has provided new opportunities to explore the emergence of these power laws, and to disentangle different sources of energy cascade. Here, we point out that recent experiments in cold atoms imply that classical turbulence is part of a larger family of scale invariant phenomena that include ideal gases. Moreover, the property of the entire family is contained in the structure of its Floquet states. For ideal gases, we show analytically that its momentum distribution acquires a $1/q^2$ tail in each dimension when it is shaken periodically. 
\end{abstract}

\maketitle

When a system is driven far away from equilibrium, energy is transferred from one length scale to another. Such processes often lead to strong fluctuations and chaotic behavior. For a classical turbulent fluid, the conventional picture is that as energy is injected at a large length scale, it cascades down to nearby smaller scales through formation of eddies, and will be finally dissipated through viscosity at a short length scale~\cite{Tur-ref,Tur-ref2,Tur-ref3}. The famous $k^{-5/3}$ scaling of the energy spectrum proposed by Kolmogorov is based on this picture~\cite{Kolmo}. However, after a century of studies, a first principle derivation of this power law is still lacking. In recent years, there have been increasing efforts to use Bose condensates to study superfluid turbulence~\cite{SF-Tur, Makoto, Onorato, Bradley, Bagnato, Shin, Marios}. Recently, Hadzibabic's group at Cambridge has shown that when a Bose condensate in a box trap is subjected to an oscillating force, its momentum distribution $n(k)$ exhibits a power law behavior~\cite{Navon}. This study has raised new issues in the research of turbulence. While conventional studies focus on velocity-velocity correlations (a four point function), momentum distribution is a two-point function, and is arguably a more fundamental quantity. 

The Cambridge experiment was explained in terms of Gross-Pitaevskii (GP) equation~\cite{Navon}.  Here, we study the power law behavior of shaken quantum gases  in a different context, not necessarily restricted to bosons. 
Our purposes are :  {\bf I} To point out that the recent experiments
~\cite{Bagnato, Shin, Marios, Navon} as well as the physics of BEC-BCS crossover~\cite{Mohit, Carlos} suggest strongly that  (A) the state of classical turbulence is part of a large ``cascade continuum" characterized by power law correlation functions, and (B) the power law behavior of the entire continuum is contained in its Floquet states that can be generated by a periodic force with a {\em single} frequency; 
{\bf II} to show explicitly that when an ideal gas in a box trap is subjected to a shaking force, its momentum distribution will acquire a $1/q^2$ power law tail, consistent with the features (A) and (B). 

{\em The cascade continuum:} 
Figure 1 shows the equilibrium phase diagram of BEC-BCS crossover of a spin-1/2 Fermi gas~\cite{Mohit, Carlos}. The horizontal axis is $-(k_{F}a_{s})^{-1}$, where $k_{F}$ is the Fermi wave vector and $a_{s}$ is the scattering length between opposite spins. 
Far on the BEC side, $0< k_{F}a_{s}\ll 1$, two fermions with opposite spin form a tightly bound bosonic molecule. The ground state is a Bose-Einstein condensate (BEC) of molecules, described by the GP equation as the Rb BEC in Ref.~\cite{Navon}. In the strongly interacting regime, $1/|k_{F}a_{s}| \leq 1$, the ground state is a BCS condensate. Recent experiments~\cite{Grimm} show that it is well described by Landau's two fluid hydrodynamics. 
It is known that the GP equation can be cast in the form of the $T=0$ two-fluid hydrodynamics Ref.~\cite{Stringari}, this means that the superfluid phase in the strongly interacting region will exhibit the same power law behavior as in Ref.~\cite{Navon} when shaken by a periodic force. 

\begin{figure}[htbp]
\vspace{10pt}
\includegraphics[angle=0,width=75mm]{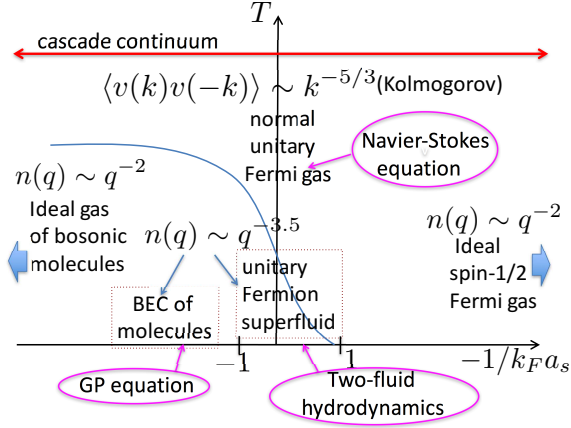}
\caption{(Color online) Different regimes of power law behavior of a Fermi gas under shaking:  The solid curve is the phase boundary separating the superfluid phase from the normal phase.  The superfluid in the region $|k_{F}a_s|^{-1} \lesssim 1$ is the unitary fermion condensate, described by Landau's two-fluid hydrodynamics.  For larger $1/k_{F}a_s$,  the system becomes a molecular condensate, described by the GP equation. Due to similar vortex dynamics of these two superfluids, a periodic force such as that in Ref.~\cite{Navon} will generate a power law in their momentum distribution. 
In the normal state near unitarity, the system is described by the Navier-Stokes equation, and will exhibit classical turbulence. 
 The region $(k_{F}a_{s})^{-1}\rightarrow  - \infty$ and $+\infty$ correspond to ideal Fermi gas and ideal Bose gas respectively. When subjected to an oscillating force, their momentum distribution will acquire a $1/q^2$ tail. 
The entire region $-\infty < (k_{F}a_{s})^{-1} < + \infty$ is likely to have power law correlations when the gas is shaken, forming a ``cascade continuum".} \label{Regime}
\end{figure}

On the other hand, it is well known that the two-fluid hydrodynamic equations reduce to  the Navier-Stokes equation in the normal state. A normal Fermi gas near unitarity must therefore exhibit turbulent behavior at sufficiently high Reynold numbers~\cite{Tur-ref,Tur-ref2,Tur-ref3}. The question is how the power law behavior of a quantum gas under shaking changes with interaction  strength $a_{s}$. Should the power law correlations persist to vanishing interaction, (which we shall show to be the case), it means that the classical 
turbulence is part of a larger family of dynamical phenomena that exhibit power law behavior, and the region of parameter space that contains this behavior 
is likely to extend over the entire interaction range $(k_{F}a_{s})^{-1}$, which we shall refer to as the ``cascade continuum".   
In the conventional studies of turbulence, the term ``cascade" means transfer of energy between different length scales through particle interactions. Here, we use this term in a more general sense to include the transfer of energy between different spatial modes caused by external forces. 

The fact that a periodic force with a single frequency $\omega$ is able to generate a power law in $n(k)$ (Ref.~\cite{Navon})  has important implications. Since the quantum evolution of periodically driven systems is given by their Floquet states, the emergence of the power law behavior in the entire cascade continuum must be reflected in its Floquet structure. Due to the complexity of the Floquet states, it is useful to disentangle the effects of interaction  from those of external forces, and to obtain exact results for the emergence of power law behavior. Therefore, we first study the shaking of an ideal gas where energy cascade is facilitated entirely by the external forces. 

{\em The shaking of an ideal gas:} 
Consider an ideal gas in a one dimensional box of length $L$ shaking with frequency $\omega$. The Schrodinger equation is 
\begin{equation}
i\partial_{t}\Psi(x,t) = H(t)\Psi(x,t), \,\,\,\, H(t)= -\partial_{x}^2 + U(x-a {\rm cos}\, \omega t)
\label{Sch} \end{equation}
where $U(x)=0$ for $|x|< L/2$,  $U(\pm L/2)=+\infty$, and $a$ is the amplitude of shaking. We have set $\hbar=1$ and $\hbar^2/2M=1$. In Supplementary Materials, we show that the shaking box in Eq.~(\ref{Sch}) is equivalent to an oscillating linear potential (used in Ref.~\cite{Navon}) through a unitary transformation, and that our results can be easily generalized to all dimensions. 

In the absence of shaking, $a=0$, $H(t)$ is the static Hamiltonian $H_{o}=-\partial_{x}^2 +U(x)$. Its eigenstates $u_{k}(x)$, referred to as ``box states", are even and odd parity states 
\begin{eqnarray}
u_{k}(x)=\sqrt{ \frac{2}{L}} {\rm cos}kx;  \,\,\, \, k= (n+1/2)(2\pi/L)
\label{even} \\
= \sqrt{ \frac{2}{L}} {\rm sin}kx \,\,\, \, k= m(2\pi/L) \hspace{0.3in} \label{odd}
\end{eqnarray}
where $n=0,1,2, ...$ and $m=1,2,3,...$.
For later discussions, we shall consider an even parity state 
$\alpha^{}_{P}(x)=\sqrt{ \frac{2}{L}} {\rm cos}Px$ with a wave vector $P$ not restricted to the set in Eq.~(\ref{even}). Its Fourier transform  
$\tilde{\alpha}^{}_{P}(q) = \int^{L/2}_{-L/2} {\rm d}x e^{-iqx} \alpha^{}_{P}(x)$ is  
\begin{equation}
\tilde{\alpha}^{}_{P}(q) = 2\sqrt{\frac{2}{L}}\left( 
\frac{ q {\rm cos}\frac{PL}{2} {\rm sin}\frac{qL}{2} - 
P  {\rm sin}\frac{PL}{2} {\rm cos}\frac{qL}{2}    }{   q^2 - P^2} \right).
\label{FT} \end{equation}
For a box state, $P=k$ in Eq.~(\ref{even}), the factor  ${\rm cos}\frac{PL}{2}$ vanishes, resulting in a  momentum distribution 
$|\tilde{\alpha}_{k}(q)|^2 \sim q^{-4}$ for large $q$. 
 The same holds for odd parity states. On the other hand, if $P$ is different from the wave vector set in Eq.~(\ref{even}), ${\rm cos}\frac{PL}{2} \neq 0$, then $|\tilde{\alpha}_{k}(q)|^2 \sim q^{-2}$. 
 Note also that in experiments, the wave vector $q$ in the momentum distribution $n(q,t)$ takes on all values,
not restricted to the discrete sets in Eq.~(\ref{even}) and (\ref{odd}).  

Since $H(t)$ is periodic in time, according to Floquet theory~\cite{Floquet1,Floquet2}, the time evolution of an initial state $\Psi_{o}(x)$ is given by : 
\begin{equation}
\Psi^{}(x,t) = \sum_{s} \langle x| \Psi^{}_{s}(t)\rangle 
\langle \Psi^{}_{s}(0)| \Psi_{o}\rangle. 
\label{Gen} \end{equation}
where $\Psi^{}_{s}(x,t)\equiv  \langle x|\Psi_{s}^{}(t)\rangle$ is the Floquet state. It satisfies Eq.~(\ref{Sch}) and is of the form
\begin{equation}
\Psi^{}_{s}(x,t)= e^{-ist} \varphi^{}_{s}(x,t), \,\,\,\, 
\varphi_{s}(x,t)=\varphi_{s}(x,t+T)
\label{Floquet} \end{equation}
where $s$ is the quasi-energy such that
$0\leq s< \omega$. 
Since $U=0$ inside the box, Eq.~(\ref{Sch}) implies that 
\begin{equation}
\varphi^{}_{s} (x) =  \sum_{\ell}
\left( A^{}_{\ell}{\rm cos}P_{\ell} x 
+ B^{}_{\ell}{\rm sin} P_{\ell}x\right), \,\,\,\,   
P_{\ell}^{2} = s+\ell \omega ,
\label{exact} \end{equation} 
where $A_{\ell}$ and $B_{\ell}$ are chosen to satisfy 
the boundary conditions for all time $t$, 
\begin{equation}
\varphi_{s}^{}(\pm L/2 - a{\rm cos}\, \omega t)= 0.
\label{BC} \end{equation}
The functions ${\rm cos}P_{\ell} x$ and 
 ${\rm sin}P_{\ell} x$ will be referred to as  ``frequency modes" of the Floquet state. The solution (Eq.~(\ref{exact})) was first constructed by Wagner~\cite{Wagner} to study the case of $\omega \ll 1/L^2$.  However, to study the energy cascade due to shaking, we  need to study some hitherto unexplored properties of Floquet states (Eq.~(\ref{exact})) for arbitrary shaking amplitudes.

As we shall see, the wave vector $P_{\ell}$ of the frequency modes must differ from the box state values in Eq.~(\ref{even}) and (\ref{odd}) for $a\neq 0$. Consequently, their Fourier transforms  behave as $1/q$ for $q\gg P_{\ell}$ as shown in Eq.~(\ref{FT}), leading to a $1/q^2$ tail in the momentum distribution. 
However, to calculate the momentum distributions of the Floqeut states (Eq.~(\ref{Floquet})) or the general states (Eq.~(\ref{Gen})), we need to understand the behaviors of the coefficients $A_{\ell}$ and $B_{\ell}$. 

We show in Supplementary Materials that Eq.~(\ref{BC}) is equivalent to the following matrix equations for  the coefficients $\{ A_{\ell}, B_{\ell} \}$ 
 \begin{equation}
 \sum_{m} W_{n,m} f_{m}=0, \,\,\,\,\, \sum_{m} \tilde{W}_{n,m} \tilde{f}_{m}=0,
\label{WW} \end{equation}
\begin{eqnarray}
W_{n,m} = \epsilon_{m} i^{m} J_{-n+m}^{}(P_{m}a){\rm cos}P_{m}^{}L/2, \,\,\,\,\, n \,\, {\rm even}  \label{AB1}\\
=\epsilon_{m}^{-1} i^{m} J_{-n+m}^{}(P_{m}a){\rm sin}P_{m}^{}L/2, \,\,\,\,\, n \,\, {\rm odd}
\label{AB2} \end{eqnarray}
\begin{eqnarray}
\tilde{W}_{n,m} = \epsilon_{m} i^{m} J_{-n+m}^{}(P_{m}a){\rm sin}P_{m}^{}L/2, \,\,\,\,\, n \,\, {\rm even}  \label{AB3}\\
=\epsilon_{m}^{-1} i^{m} J_{-n+m}^{}(P_{m}a){\rm cos}P_{m}^{}L/2, \,\,\,\,\, n \,\, {\rm odd}
\label{AB4} \end{eqnarray}
 where $\epsilon_{m}=1$ and $i$ for even and odd $m$;  $J_{n}(x)$ is the Bessel function; $f_{m}=A_{m}$ and $B_{m}$ ( and $\tilde{f}_{m}=B_{m}$ and $A_{m}$) 
 for even and odd $m$, respectively. The two equations in Eq.~(\ref{WW}) are for the sets $(... B_{-1}, A_{0}, B_{1}, A_{2}, ..)$ and $(... A_{-1}, B_{0}, A_{1}, B_{2}, ..)$. The quasi-energy $s$ enters the matrices $W$ and $\tilde{W}$ through $P_{m}= m\omega + s$, Eq.~(\ref{exact}). It is determined by setting ${\rm Det}||W||={\rm Det}||\tilde{W}||=0$.  We have found the solutions of Eq.~(\ref{WW})  and have verified their validity by solving Eq.~(\ref{Sch}) using different methods.
 (See Supplementary Materials). The properties of  Floquet states are as follows.

{\em 1. Floquet families :}  
As $a\rightarrow 0$, the Floquet state $\Psi_{s}(x,t)$ must reduce to one of the box states, say,  $e^{-ik^2 t}u_{k}(x)$, which can be written as $e^{-is_{k}t}\left[ e^{-i\ell_{k}\omega t}u_{k}(x) \right]$, 
\begin{equation}
\ell_{k}={\rm Int}\left(\frac{k^2}{\omega} \right), \,\,\,\, \frac{s^{}_{k}}{\omega} = \{  \frac{k^2}{\omega} \} 
\end{equation}
where ${\rm Int}(x)$ and $\{ x \}>0$ are the integer and fractional part of $x$, $x = {\rm Int}(x) + \{ x\}$.
In other words, the box state $e^{-ik^2 t}u_{k}(x)$ is a Floquet state with quasi-energy $s_{k}$ and a single frequency mode at $\ell_{k}$. This can also be seen from Eqs.~(\ref{AB1})-(\ref{AB4}). When $a=0$, only the Bessel functions with $m=n$ survive, which means either 
${\rm cos}(P_{n}L/2)$ or ${\rm sin}(P_{n}L/2)=0$, i.e. $P_{n}$ must be the wave vector of a box state (Eqs.~(\ref{even}) and (\ref{odd})). 

\begin{figure}[htbp]
\vspace{10pt}
\includegraphics[angle=0,width=70mm]{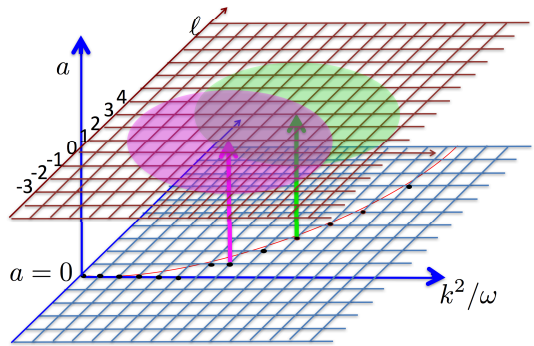}
\caption{(Color online)
Schematic representation of the structure of the Floquet states in the space of frequency and spatial mode, ($\ell-k^2$), with $k^2$ measured in units of $\omega$. The solid curve is the energy $k^2$. 
At $a=0$, each Floquet state is a box state $u_{k}$ with a single frequency mode $\ell_{k}$, corresponding to a point (a black dot) in the $\ell-k^2$ plane.   
Its quasi-energy $s_{k}$ is the difference between the height of the curve and the location of the dot below, $s_{k}= k^2 - \ell_{k} \omega$. 
As $a$ increases, the number of frequency  modes proliferates, and the quasi-energy becomes $s_{k,a}$.  Since the wave vectors $P_{\ell}$ of these frequency modes differ from those of box states, each of them is composed of a large number box states $\{ u_{p}\}$ with distribution 
$\tilde{\alpha}_{p}(P_{\ell})$. The state $s_{k,a}$ then 
occupies a region in the $\ell-k^2$ plane. The two circular discs depicted represent two Floquet states at $a\neq 0$ grown out from two different box state at $a=0$. 
The family of states $s_{k,a}$ for all $a$ will be referred to as the Floquet family rooted at $k$. Among all box states, $|s_{k,a}\rangle$ has largest overlap with its root state. 
}
\label{root}
\end{figure}

For non-zero $a$, none of  the Bessel functions in Eqs.~(\ref{AB1})-(\ref{AB4}) vanish. Two processes then occur.  (i) A box state, say, at $k_{o}$ (i.e. a Floquet state with a single frequency mode at $\ell_{k_{o}}$),   will cause other frequency modes to grow. (ii) At the same time, its wave vector will shift away from $k_{o}$. Both features  are results of  energy cascade. The former is a result of the coupling between frequency modes differing by energy $\sim \omega$. The latter is the result of energy transfer from $k_{o}$ to nearby box states, as ${\rm cos}Px$ with $P$ close to a box state $k_{o}$ is made up of many box states near $k_{o}$ with distribution $\tilde{\alpha}_{k_{o}}(P)$.  (See also Section III of Supplementary Materials). 
The family of Floquet states that emerges from the box state $k_{o}$ as $a$ increases will be referred to as the Floquet family ``rooted" at $k_{o}$, with  quasi-energy 
 $s=s_{k_{o}, a}^{}$, ($s_{k_{o}, a=0}^{}=s_{k_{o}}^{}$). The general structures of these families are shown schematically in Fig. \ref{root}.

Consequently, we can write the Fourier transform of the state in Eq.~(\ref{Gen}) as 
\begin{equation}
\tilde{\Psi}^{}(q,t) = \sum_{k} e^{-is_{k,a}^{}t} \langle q| s_{k,a}^{}; t\rangle 
\langle s_{k,a}^{}; 0| \Psi_{o}\rangle. 
\label{Psiq} \end{equation}
where we have use the simplified  notation $|\varphi^{}_{s_{k,a}}; t\rangle \equiv 
|s_{k,a}; t\rangle$. 
The momentum distribution of $\tilde{\Psi}^{}(q,t)$ at integer period is $n(q)\equiv n(q,NT)
\equiv |\tilde{\Psi}^{}(q,NT)|^2$, or
\begin{eqnarray}
n(q)= 
\sum_{k}|G_{k}(q)|^2  
+\sum_{k\neq k'} G^{\ast}_{k'}(q)G^{}_{k}(q)e^{iNT(s_{k',a}^{}-s_{k,a}^{})}  \hspace{0.3in}\label{nq} \\
G_{k}(q)=\langle q | s_{k,a}^{}; 0\rangle \langle  s_{k,a}^{}; 0| \Psi_{o}\rangle.  \hspace{1.0in}  \end{eqnarray}
The second term in Eq.~(\ref{nq}) represents the dephasing between different Floquet states. 
It will decrease in strength for large time $NT$. If this term decreases to zero, $n(q)$ will be given by the first term, which is the sum of the momentum distribution of the Floquet state that overlaps strongly with the initial state. 

\begin{figure}[htbp]
\includegraphics[angle=0,width=70mm]{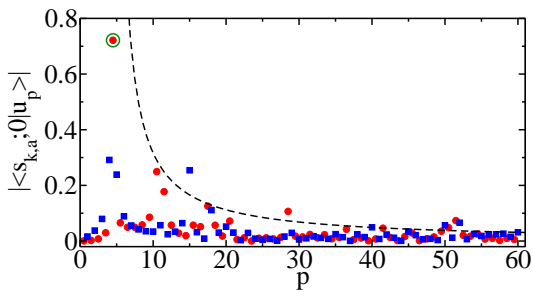}
\caption{(Color online) The magnitude of the overlap $|\langle  s_{k,a}^{}; 0| u_{p}\rangle|$ versus $p$: for $k=4.5(2\pi/L)$, $\omega=100(2\pi/L)^2$, $L=1$, and $a\omega L= 76$. (See Supplementary Materials). Squares and circles correspond to box states $u_{p}$ with even and odd parity. The circled peak is the root state.}
\label{fig_coeff_s}
\end{figure}

\begin{figure}[htbp]
\includegraphics[angle=0, width=70mm]{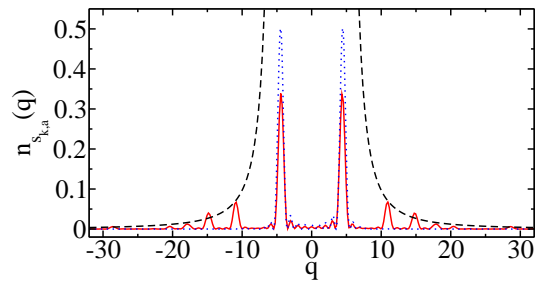}
\caption{(Color online) The red solid line shows the momentum distribution $n_{s_{k,a}}(q)$ for the Floquet state 
shown in Figure \ref{fig_coeff_s} at integer period. The
blue dotted line shows the momentum distribution of the root state at $k= 4.5$. The dash line shows a $1/q^2$ tail.}
\label{fig_dis_single}
\end{figure}

{\em 2. Overlap and momentum distribution:} To calculate the overlap $ \langle  s_{k,a}^{}; 0| \Psi_{o}\rangle$ in Eq.(\ref{Psiq}), it is sufficient to calculate 
$ \langle  s_{k,a}^{}; 0| u_{p}\rangle$, which can be obtained from the solution of Eq.~(\ref{AB1}) to (\ref{AB4}).  
In Fig. \ref{fig_coeff_s}, we have plotted the magnitude of this overlap 
as a function of $p$ for a shaking amplitude $a\gg 1/\omega L$. In terms of un-scaled length and time, it means the momentum due to the shaking $Ma\omega$ is much larger than the confinement momentum $\hbar/L$. Figure \ref{fig_coeff_s}  shows that the overlap is maximum at  $p=k$, i.e. the overlap is the largest if the box state $|u_{p}\rangle$ is the root of the Floquet family $s_{k,a}$. Figure \ref{fig_coeff_s} also shows a sequence of larger discrete peaks separated from 
neighboring ones with energy difference $\sim \omega$,  as well as other smaller peaks surrounding these larger peaks. They are caused by the cascade processes (i) and (ii) discussed previously. Despite the somewhat random appearance of these peaks, it actually has a  regularity structure as seen from their momentum distribution 
$n_{s_{k,a}}(q,t)\equiv |\langle q|s_{k,a}; t\rangle|^2 $ at integer period $t=NT$, 
as shown in Figure \ref{fig_dis_single}. 
It has a sequence of pronounced peaks separated from
 neighboring ones with energy $\omega$. This means there is still 
considerable phase coherence in the Floquet state.  Finally, due to the $1/q$ tail of the Fourier transform of each frequency mode,  $n_{s_{k,a}}(q)$ acquires a $1/q^2$ tail.

\begin{figure}[htbp]
\vspace{20pt}
\includegraphics[angle=0,width=70mm]{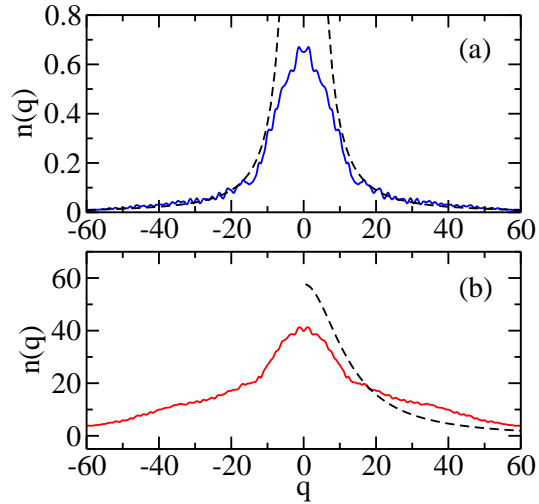}
\caption{(Color online)  Upper panel: the momentum distribution of  a Fermi gas evolved from a zero temperature Fermi distribution averaged over a periodic after long time $t=NT$, $N\gg 1$, with parameters same as Figure \ref{fig_coeff_s}. 
The Fermi wave vector of the initial state is $k_{F}=8$.
Lower panel: The corresponding momentum distribution of a Bose gas evolved from at low temperature equilibrium distribution. The temperature and chemical potential  (in units of $(2\pi/L)^2$) of the initial state is $k_B T=9000$ and $\mu=-154.5$. The dashed line is the equilibrium Bose distribution function.}
\label{fig_fermi_bose}
\end{figure}

Equations~(\ref{Gen}) and (\ref{Psiq}) can be easily generalized to the evolution of density matrices.  For an ideal Fermi gas and Bose gas with an initial equilibrium distribution, 
their  momentum distributions averaged over a period after long time shaking is 
are shown in Fig.~\ref{fig_fermi_bose}.  Indeed we have found the second term in Eq.~(\ref{nq}) vanishes. 
A $1/q^2$ tail is apparent in the case of Fermi gas when $q>k_{F}$, since all the Floquet states rooted within the Fermi momentum $k_{F}$ all have a $1/q^2$ tail. 
In contrast, the Bose distribution does not have a sharp cut off in energy  space, the Floquet states rooted  below and above $q$ all 
contribute to the momentum distribution at $q$. As a result, $n(q,t)$ does not have a clean $1/q^2$ tail. Instead, there is a strong depletion of occupation at small momentum. Finally, we note that generalizing to 3D, the momentum distribution $n(q_x, q_y, q_z; t) $ will behave as $(q_{x}q_{y} q_{z})^{-2}$ for large ${\bf q}$.

{\em Final remarks:} The $1/q^2$ tail in the momentum distribution is caused by the shift of wave vector $P_{\ell}$ of the frequency modes from its original quantized value  in a static box. Such shifts are required to satisfy the boundary condition once other frequency modes appear. These shifts can also be viewed as the appearance of an effective ``energy dependent"  box size. This feature should also persist in the presence of interaction, even though interaction will lead to a different power law.

We thank Nir Navon for communications of his experiment and for comments on the first draft of this paper; Jayaprakash for stimulating discussions on turbulence. TLH acknowledges the support by NSF Grant DMR1309615, MURI Grant FP054294-D, and NASA Fundamental Physics Grant 1518233.

\bibliographystyle{apsrev4-1}
\nocite{apsrev41Control}
\bibliography{mybib}

\end{document}


\title{Supplemental Materials \\ Energy Cascade in Quantum Gases}
\author{Tin-Lun Ho and X. Y. Yin}
\affiliation{Department of Physics, The Ohio State University, Columbus, OH 43210, USA}
\date{\today}

\maketitle
\onecolumngrid
\renewcommand\theequation{S\arabic{equation}}


\section{{\bf (I)}  Equivalence between a shaking potential and an oscillating linear potential}

The Schr\"odinger equation of a particle in a trap $U(x)$ subjected to an oscillating linear potential is
\begin{equation}
i\hbar\partial_{t} |\eta(t)\rangle = \left( \frac{p^2}{2M}  + U(x) - \gamma \frac{x}{L} {\rm cos}\omega t \right) |\eta(t)\rangle.
\label{1A} \end{equation}
Such oscillating potential is used in the recent Cambridge experiment~\cite{Navon}. 
Through a unitary transformation
\begin{equation} 
|\Psi(t)\rangle = e^{-ip a(t)/\hbar} e^{iM\dot{a} x/\hbar} e^{i\int^{t}_{0} M\dot{a}(t')^2/2 dt' /\hbar}|\eta (t)\rangle,  \,\,\,\,\,\,\,  a(t) = a{\rm cos}\omega t, \,\,\,\,\,\,
\gamma/L = Ma \omega^2
\label{UT} \end{equation}
or 
\begin{equation} 
|\Psi(t)\rangle = {\rm exp}\left[ -\frac{ip a{\rm cos}\, \omega t}{\hbar}\right] 
{\rm exp}\left[-\frac{iMa \omega {\rm sin}\omega t}{\hbar} \right]
{\rm exp}\left[ \left(\frac{iM\omega^2 a^2}{2\hbar}\right)\left(t - \frac{{\rm sin}2\omega t}{2 \omega}  \right) \right]
|\eta (t)\rangle, 
\end{equation}
Equation~(\ref{1A}) can be written as
\begin{equation}
i\hbar\partial_{t} |\Psi(t)\rangle = \left( \frac{p^2}{2M}  + U(x-a{\rm cos}\omega t ) \right) |\Psi(t)\rangle. 
\end{equation}
which is the Schr\"odinger equation of a particle in  a shaking potential with shaking amplitude $a$. If we denote the momentum distribution of $|\Psi(t)\rangle$ and $|\eta(t)\rangle$ as 
\begin{equation}
n_{\eta}(q,t) = |\langle q|\eta(t)\rangle|^2, \,\,\,\,\,\, {\rm and} \,\,\,\,\,\,
n_{\Psi}(q,t) = |\langle q|\Psi(t)\rangle|^2, 
\end{equation}
it is straightforward to show that at integer time $t=NT$, $T=2\pi/\omega$, we have 
\begin{equation}
n_{\eta}(q,NT) = n_{\Psi}(q,NT) 
\end{equation}

This result can be generalized to higher dimensions. For example, in 3D, an oscillating force in rectangular box with length $L_{x}, L_{y}, L_{z}$ is simply 
\begin{equation}
i\hbar\partial_{t} |\eta(t)\rangle = \sum_{i=x,y,z}\left( \frac{p_{i}^2}{2M}  + U(x_{i}) - \gamma_{i} \frac{x_{i}}{L_{i}} {\rm cos}\omega t \right) |\eta(t)\rangle.
\label{1AA} \end{equation}
which is separable into three separate directions. 
In our calculations that generate Fig. 3 and 4, we have considered the case $Ma\omega \gg \hbar/L$. From Eq.~(\ref{UT}), it   
corresponds to $\gamma/(\hbar \omega) \gg 1$.

\section{{\bf (II)}  Derivation and solution of Eqs.~(10)-(13) : }

The boundary condition Eq.~(8),
\begin{equation}
\sum_{\ell}e^{-i\ell \omega t} \left( A_{\ell}{\rm cos}\left(P_{\ell}\left(\pm \frac{L}{2}- a {\rm cos}\omega t\right)\right) + B_{\ell}{\rm sin}\left(P_{\ell}\left(\pm \frac{L}{2}- a {\rm cos}\omega t\right)\right) \right)=0
\label{BBCC} \end{equation}
can be written as 
\begin{equation}
\sum_{\ell}e^{-i\ell \omega t} \left( 
\left[ A_{\ell}{\rm cos}\left(\frac{P_{\ell} L}{2} \right) \pm B_{\ell}{\rm sin}\left(\frac{P_{\ell} L}{2}\right)\right]
{\rm cos}(P_{\ell}a {\rm cos}\omega t) + 
\left[ \pm A_{\ell}{\rm sin}\left( \frac{P_{\ell} L}{2}\right) 
- B_{\ell}{\rm cos}\left(\frac{P_{\ell} L}{2}\right)\right]
{\rm sin}(P_{\ell}a {\rm cos}\omega t) \right) =0. 
\label{BC1} \end{equation}
Using the fact that 
\begin{eqnarray}
{\rm cos}(z{\rm cos}\phi) & = \sum_{m} (-1)^{m} J_{2m}(z) e^{2im\phi} \\
{\rm sin}(z{\rm cos}\phi) & = \sum_{m} (-1)^{m} J_{2m+1}(z) e^{i(2m+1)\phi}
\end{eqnarray}
we can write Eq.~(\ref{BC1}) as 
\begin{eqnarray}
\sum_{\ell}e^{-i\ell \omega t} 
\left[ A_{\ell +2m}{\rm cos}(P_{\ell +2m} L/2) \pm B_{\ell+2m}{\rm sin}(P_{\ell+2m} L/2)  \right](-1)^m J_{2m}(P_{\ell + 2m}a) \hspace{0.5in} \nonumber \\
+ \sum_{\ell}e^{-i\ell \omega t} 
\left[ \pm A_{\ell +2m +1 }{\rm sin}(P_{\ell +2m +1} L/2) - B_{\ell+2m+1}{\rm cos}(P_{\ell+2m+1} L/2)  \right](-1)^m J_{2m+1}(P_{\ell + 2m+1}a) =0
\label{BC2} \end{eqnarray}
Due to the $\pm$ sign, Eq.~(\ref{BC2}) consists of two equations. Adding and subtracting these two equations, and setting all the coefficients of the phase factors to zero, we have 
\begin{equation}
\sum_{m}(-1)^m \left[  A_{\ell + 2m} {\rm cos}(P_{\ell+2m}L/2) J_{2m}(P_{\ell+2m}a)
-  B_{\ell + 2m+1} {\rm cos}(P_{\ell+2m+1}L/2) J_{2m+1}(P_{\ell+2m+1}a)\right] =0
\label{1st}
\end{equation}
and
\begin{equation}
\sum_{m}(-1)^m \left[  B_{\ell + 2m} {\rm sin}(P_{\ell+2m}L/2) J_{2m}(P_{\ell+2m}a)
-  A_{\ell + 2m+1} {\rm sin}(P_{\ell+2m+1}L/2) J_{2m+1}(P_{\ell+2m+1}a)\right] =0.
\label{2nd}
\end{equation}

--------------

Let us consider Eq.~(\ref{1st}) with even $\ell=2n$, and Eq.~(\ref{2nd}) with odd $\ell=2n-1$.  Equations~(\ref{1st}) and (\ref{2nd}) then become
\begin{equation}
\sum_{m}(-1)^m \left[  A_{2m} {\rm cos}(P_{2m}L/2) J_{2m-2n}(P_{2m}a)
-  B_{2m+1} {\rm cos}(P_{2m+1}L/2) J_{2m-2n+1}(P_{2m+1}a)\right] =0
\label{1st1}
\end{equation}
and
\begin{equation}
\sum_{m}(-1)^m \left[  B_{2m-1} {\rm sin}(P_{2m-1}L/2) J_{2m-2n}(P_{2m-1}a)
-  A_{2m} {\rm sin}(P_{2m}L/2) J_{2m-2n+1}(P_{2m}a)\right] =0. 
\label{2nd1} 
\end{equation}
Equations~(\ref{1st1}) and (\ref{2nd1}) are the equations for the sequence $(...B_{-1}, A_{0}, B_{1}, A_{2}, ...)$.  Equations~(\ref{1st1}) can be written as 
\begin{equation}
\sum_{m}\left(J_{-2n +m}(P_{m}a){\rm cos}\left(\frac{P_{m}L}{2}\right) i^{m} \epsilon_{m}
\right)f_{m} =0,
\end{equation}
where $f_{m}= A_{m}$ for even $m$, and $f_{m}= B_{m}$ for odd $m$; and 
\begin{equation}
\epsilon_{m}=1  \,\,\,\,\, {\rm for }\,\,\,\, {\rm even } \,\, m ; \,\,\,\,\,
\epsilon_{m}=i  \,\,\,\,\, {\rm for }\,\,\,\, {\rm odd } \,\, m.
\label{1st2}\end{equation}
Similarly, Eq.~(\ref{2nd1}) can be written as 
\begin{equation}
\sum_{m}\left(J_{-2n+1 +m}(P_{m}a){\rm sin}\left(\frac{P_{m}L}{2}\right) i^{m} \epsilon^{-1}_{m}
\right)f_{m} =0,
\label{2nd2}\end{equation}
Equations~(\ref{1st2}) and Eq.~(\ref{2nd2}) can be summarized in a matrix equation
\begin{equation}
\sum_{m}W_{n,m}f_{m}=0
\label{wf1} \end{equation}
\begin{eqnarray}
W_{n,m}= \epsilon_{m} i^{m} J_{-n+m}(P_{m}a) {\rm cos}(P_{m}L/2) , \,\,\,\,\, {\rm for} \,\,\, n \,\,{\rm even}    \\
W_{n,m}= \epsilon^{-1}_{m} i^{m} J_{-n+m}(P_{m}a) {\rm sin}(P_{m}L/2) , \,\,\,\,\, {\rm for} \,\,\, n \,\,{\rm odd}
\end{eqnarray}

------------------

We now consider Eq.~(\ref{1st}) with odd $\ell=2n-1$, and Eq.~(\ref{2nd}) with even $\ell=2n$.  Eqs.~(\ref{1st}) and (\ref{2nd}) then become
\begin{equation}
\sum_{m}(-1)^m \left[  A_{2m-1} {\rm cos}(P_{2m-1}L/2) J_{2m-2n}(P_{2m-1}a)
-  B_{2m} {\rm cos}(P_{2m}L/2) J_{2m-2n+1}(P_{2m}a)\right] =0 , 
\label{new1st1}\end{equation}
\begin{equation}
\sum_{m}(-1)^m \left[  B_{2m} {\rm sin}(P_{2m}L/2) J_{2m-2n}(P_{2m}a)
-  A_{2m+1} {\rm sin}(P_{2m+1}L/2) J_{2m-2n+1}(P_{2m+1}a)\right] =0. 
\label{new2nd1} \end{equation}
Equations~(\ref{new1st1}) and (\ref{new2nd1}) are the equations for the sequence $(...A_{-1}, B_{0}, A_{1}, B_{2}, ...)$.
 Equation~(\ref{new1st1}) can be written as 
\begin{equation}
\sum_{m}\left(J_{-2n +m}(P_{m}a){\rm sin}\left(\frac{P_{m}L}{2}\right) i^{m} \epsilon_{m}
\right)\tilde{f}_{m} =0,
\label{new1st2}\end{equation}
where $\tilde{f}_{m}= B_{m}$ for even $m$, and $\tilde{f}_{m}= A_{m}$ for odd $m$.
Similarly, Eq.~(\ref{new2nd1}) can be written as 
\begin{equation}
\sum_{m}\left(J_{-2n+1 +m}(P_{m}a){\rm cos}\left(\frac{P_{m}L}{2}\right) i^{m} \epsilon^{-1}_{m}
\right)\tilde{f}_{m} =0,
\label{new2nd2}\end{equation}
Equations~(\ref{new1st2}) and Eq.~(\ref{new2nd2}) can be summarized in a matrix equation
\begin{equation}
\sum_{m}\tilde{W}_{n,m}\tilde{f}_{m}=0
\label{wf2}\end{equation}
\begin{eqnarray}
\tilde{W}_{n,m}= \epsilon_{m} i^{m} J_{-n+m}(P_{m}a) {\rm sin}(P_{m}L/2) , \,\,\,\,\, {\rm for} \,\,\, n \,\,{\rm even}    \\
\tilde{W}_{n,m}= \epsilon^{-1}_{m} i^{m} J_{-n+m}(P_{m}a) {\rm cos}(P_{m}L/2) , \,\,\,\,\, {\rm for} \,\,\, n \,\,{\rm odd}. 
\end{eqnarray}

To solve Eq.~(\ref{wf1}) and Eq.~(\ref{wf2}), we truncate them into a large square matrix. For each Floquet family $s_{k,a}$,  the range of momentum 
$[p_{min}, p_{max}]$ and the range of frequency  $[\ell_{min}, \ell_{max}]$  are chosen so that the root state with wavevector $k$ and frequency model $\ell_{k}$ is at the center of the interval. This procedure is justified as long as the frequency modes and the wavevector range of the Floquet state are well contained within the interval. We have also verified our solutions with a separate calculation described in the next section.


\section{{\bf (III)}  Alternative method of finding the Floquet states of Eq.~(1) : }
We have also solved the Schr\"odinger Eq.~(1) in a different way. This method also provides different insights. Note that Eq.~(1) can be recast as
\begin{equation}
i \partial_{t}|\Phi(t)\rangle= \left(-\partial_{x}^2 + U(x) + a \omega {\rm sin}\omega t \hat{p} \right) |\Phi(t)\rangle
\label{2Sch} \end{equation}
where $p\equiv - i \partial_{x}$ and 
\begin{equation}
|\Psi(t)\rangle = e^{- i a \hat{p} {\rm cos} \omega t} |\Phi(t)\rangle. 
\end{equation}
The relation between the solution $\Psi(x,t)$ [Eq.~((7))] and $\Phi(x,t)$ is 
\begin{equation}
e^{-ist} \sum_{\ell}  e^{-i\ell \omega t} \left( A_{\ell} {\rm cos}P_{\ell}x + 
B_{\ell} {\rm sin}P_{\ell} x \right) = 
e^{-ist} \sum_{\ell'}  e^{-i\ell' \omega t} D_{\ell', k}u_{k}(x-a{\rm cos}\omega t).
\end{equation}
Since $\langle q| \Psi(t)\rangle = e^{-iaq{\rm cos}\omega t} \langle q| \Phi(t)\rangle$
The momentum distribution of $|\Psi(t)\rangle$ and $|\Phi(t)\rangle$ are identical,  
\begin{equation}
|\langle q| \Psi(t)\rangle|^2 = |\langle q| \Phi(t)\rangle|^2.
\end{equation} 
To find the Floquet state $|\Phi_{s}(t)\rangle$ of Eq.~(\ref{2Sch}), we expand it as 
\begin{equation}
\Phi_{s}^{}(x,t)=e^{-ist}\sum_{\ell}e^{-i\ell \omega t} \Phi_{\ell}(x) \equiv  e^{-ist}
\sum_{\ell}e^{-i\ell \omega t} D_{\ell, k}u_{k}(x),
\end{equation}
where $u_{k}(x)$ are the box states. Eq.~(\ref{2Sch}) can then be written as 

\begin{equation}
(s+ \ell \omega - k^2) D_{\ell, k} = \sum_{k'} Q_{k,k'}(D_{\ell+1, k'}- D_{\ell-1, k'})
\label{2F} \end{equation}

\begin{equation}
Q_{k,k'}= i (a\omega/2) \langle u_{k}|p|u_{k'}\rangle. 
\end{equation}
If $u_{k}$ and $u_{k'}$ are of even and odd parity, we have
\begin{equation}
\langle u_{k}|p|u_{k'}\rangle 
= \frac{2}{L}\int^{L/2}_{-L/2} {\rm cos}kx \frac{\partial_{x}}{i} {\rm sin} k'x {\rm d} x
\end{equation}
and we have 
\begin{equation}
Q_{k,k'}= \frac{a \omega}{L} \frac{2k k'}{k^2 - k'^{2}}\left( {\rm sin} \frac{kL}{2} {\rm cos}\frac{k'L}{2}\right)
\label{Q} \end{equation}
The quantity in the bracket is  either $+1$ or $-1$, since $k=(2\pi/L)(n+1/2)$, and $k'=m(2\pi/L)$. 

\vspace{0.2in}

We have solved Eq.~(\ref{2F}) numerically, and have obtained the same momentum distribution as the method in the previous section. Eq.~(\ref{2F}), however, offers a different view of the energy cascade process. Eq.~(\ref{2F}) shows that a mode $D_{\ell, k}$ with frequency $\ell$ and  box momentum $k$ will be coupled strongly to the mode  $D_{\ell \pm 1, k'}$ if the energy difference $k^2-k'^2$ is close to $\omega$.  This is the mechanism for  generating new frequency modes, and the coupling is between modes of different parity.  This is the process (i) discussed in the text. On the other hand, the matrix element $Q$ is very large for nearby states $k$ and $k'$ with different parity. Even though the energy difference between $k^2$ and $k'^2$ is far from $\omega$, these processes as well as their higher order scattering, will lead to growth of $D_{\ell, k'}$ for $k'$ near $k$. This is the process (ii) discussed in the text.